\documentclass[12pt]{article}

\usepackage{graphicx}
\usepackage{hyperref}
\usepackage{a4wide}
\usepackage{amsmath}
\usepackage{amssymb}
\newcommand{\MeV}{\,\text{MeV}}
\newcommand{\GeV}{\,\text{GeV}}
\newcommand{\TeV}{\,\text{TeV}}
\newcommand{\jet}{\text{jet}}
\newcommand{\chg}{\text{chg}}
\newcommand{\tower}{\text{tower}}

\newcommand{\order}[1]{{\cal O}\left(#1\right)}

\date{}
\title{\sf Fluctuations and asymmetric jet events\\
    in Pb\,Pb collisions at the LHC}
\author{\sf
  Matteo Cacciari,\!$^{1,2}$ Gavin P.~Salam$^{3,4,1}$ and Gregory Soyez$^{5}$ \\
\\
{\sf\small $^1$LPTHE, UPMC Univ.~Paris 6 and CNRS UMR 7589, Paris, France}\\
{\sf\small $^2$Universit\'e Paris Diderot, Paris, France}\\
{\sf\small $^3$CERN, Department of Physics, Theory Unit, CH-1211 Geneva 23, Switzerland}\\
{\sf\small $^4$Department of Physics, Princeton University, Princeton, NJ 08544, USA}\\
{\sf\small $^5$Institut de Physique Th\'eorique, CEA Saclay, CNRS URA 2306, F-91191 Gif-sur-Yvette, France}
}

\begin{document}

\maketitle

\begin{abstract}
  Recent LHC results concerning full jet-quenching in Pb\,Pb
  collisions have been presented in terms of a jet asymmetry
  parameter, measuring the imbalance between the transverse momenta of
  leading and subleading jets.
  We examine the potential sensitivity of this distribution to
  fluctuations from the heavy-ion background. 
  Our results suggest that a detailed estimate of the true
  fluctuations would be of benefit in extracting quantitative
  information about jet quenching.
  We also find that the apparent impact of fluctuations on the jet
  asymmetry distribution can depend significantly on the choice of
  low-$p_t$ threshold used for the simulation of the hard $pp$ events.
\end{abstract}

In the quest to understand the properties of the medium generated in
high-energy heavy-ion collisions, the past decade has seen extensive
study of medium-induced modifications to the production of high
transverse momentum objects~\cite{reviews}.
It has been conclusively established at RHIC that the spectra of
high-momentum hadrons are significantly suppressed, by a factor of
$R_{AA} \simeq 0.2$ relative to the appropriate rescaling of the $pp$
spectra.
This effect is generally attributed to their (or their originating
parton's) interaction with the medium.

Recently, significant attention has been directed to jets.
Compared to hadrons, jets are interesting because, at least in $pp$
collisions, there is a closer, and perturbatively quantifiable,
connection between a jet's momentum and that of its originating
parton.
STAR~\cite{STAR} and PHENIX~\cite{Lai:2009ai} have presented first
(preliminary) measurements of full jets produced in AuAu collisions
with transverse momenta in the $20-40\GeV$ range and found that jet
spectra are also suppressed, though by a potentially more modest
factor than for hadrons.
Recently, ATLAS~\cite{ATLAS:2010bu} has published studies of the
\emph{correlations} between the momenta of the two leading jets, with
the striking observation that a significant fraction of events show a
strong imbalance between the $p_t$'s of the leading jet and the first
subleading jet on the opposite side of the event. CMS has shown
similar preliminary results in
Ref.~\cite{CMSTalk}\footnote{Subsequently published in
  \cite{Collaboration:2011sx}.} and first phenomenological
interpretations have been given in Ref.~\cite{Urs+Co}.

\begin{figure}
  \centering
  \includegraphics[height=0.7\textwidth,angle=90]{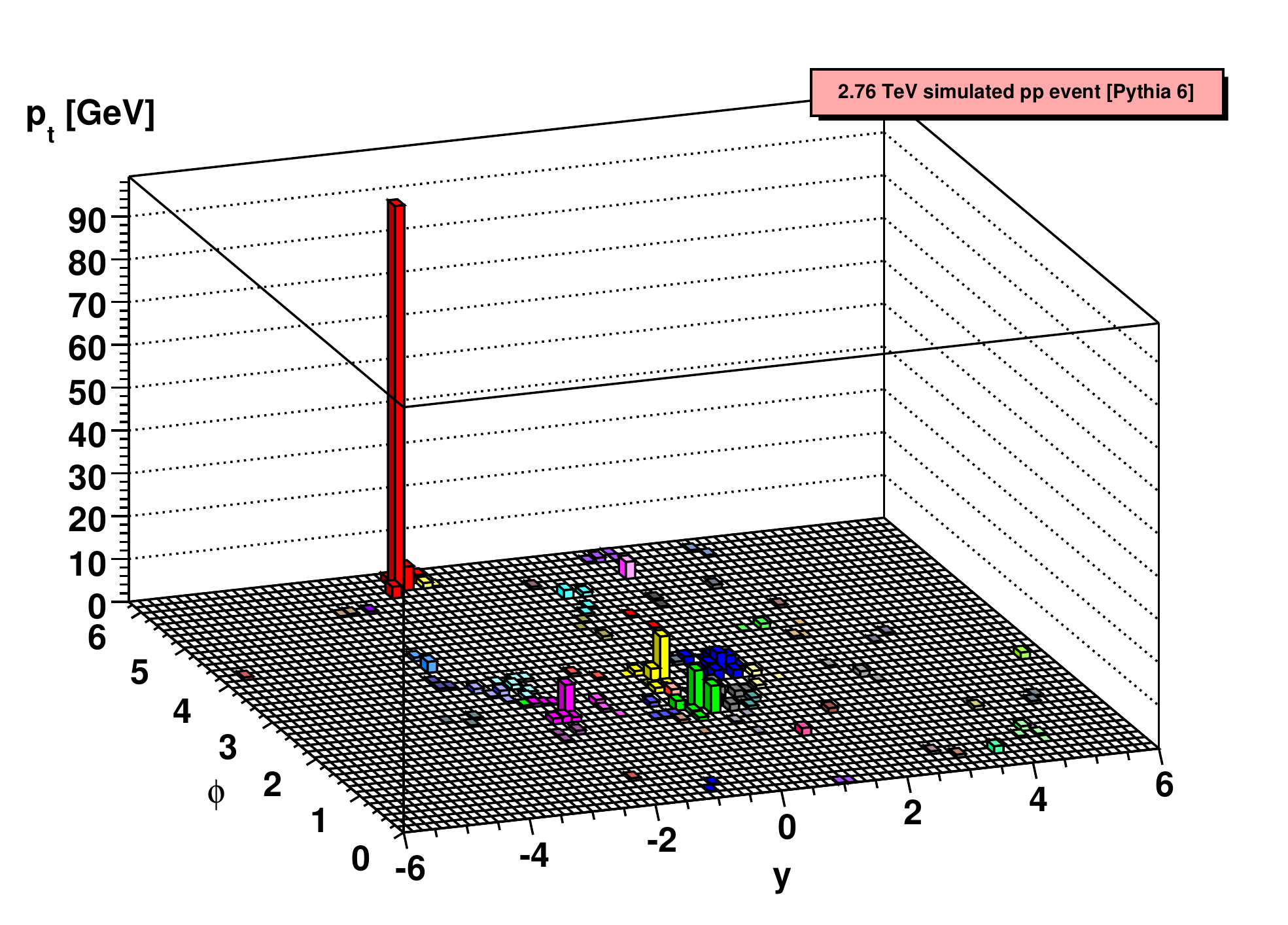}
  \caption{A simulated $pp$ event from Pythia 6.423 (centre-of-mass
    energy $\sqrt{s} = 2.76\TeV$; the missing transverse energy was
    zero).
    We find that for 1 in every 300 events with a jet with
    $p_{t1} > 100\GeV$, the second hardest jet has $p_{t2} <
    p_{t1}/3$. A more accurate estimation of this number would benefit
    from the combination of 2-jet, 3-jet, 4-jet, \ldots\ samples, using
    for example one of the multijet matching methods reviewed
    in~\cite{Alwall:2007fs}. 
  }
  \label{fig:pythia-event}
\end{figure}
Dijet imbalances can occur also in normal $pp$ events, due to emission
of multiple gluons (cf.\ the simulated Pythia event shown in
fig.~\ref{fig:pythia-event}), but they are quite rare.
To quantify how much more often they arise in Pb\,Pb collisions, ATLAS
and CMS have shown distributions of the jet asymmetry $A_J$
\begin{equation}
  \label{eq:AJ}
  A_J = \frac{E_{T1} - E_{T2}}{E_{T1} + E_{T2}}\,,
\end{equation}
expressed in terms of the transverse energies of the leading and
subleading jets, respectively $E_{T1}$ and $E_{T2}$.
The main quantitative evidence for jet quenching comes in the form of
a significant enhancement of the asymmetry in the region around
$A_J\simeq 0.4$ (fig. 3 of \cite{ATLAS:2010bu} and p. 26 of \cite{CMSTalk}).

In extracting the distribution of $A_J$, the experiments must contend
with the fact that each jet may be contaminated with
$\order{100-150\GeV}$ of transverse momentum from the medium
particles, usually referred to as the background.\footnote{The
  distinction between medium particles and jet particles is not
  necessarily very legitimate physically, however it may still make sense
  to think of an expected level of background transverse momentum.}
To calculate the $A_J$ for a given dijet event, each jet's momentum is
corrected for the expected level of background activity in the jet,
usually 
estimated from the activity  elsewhere in the
event, preferably at similar rapidities 
(see e.g. \cite{Kodolova:2007hd,Cacciari:2010te,ATLAS:2010bu}).
Such a correction cannot, however, account for the fact that the
background fluctuates from point to point within the event (even at
the same rapidity), so that the momentum subtracted from the jet will
inevitably differ from the background actually present in the jet; nor
does it account for fluctuations in the detector's response to the
background and jet particles.

Fluctuations are of course a common issue for jet measurements even in $pp$
collisions, notably due to randomness in the response of calorimeters.
However two novelties may be relevant concerning fluctuations for
heavy-ion collisions. Firstly the LHC heavy-ion medium has only just
been produced and it is probably fair to assume that its
fluctuations\footnote{Including their standard deviation, correlations
  from point to point within the event, non-Gaussianities, etc.}  are
less well understood than those of the detectors, which have been the
object of study for many years. Secondly, the absolute size (i.e. in GeV) of
detector fluctuations scales roughly as the square-root of the jet
energy, meaning that they are less important for low-$p_t$ jets than
for high-$p_t$ jets, whereas
background fluctuations are probably largely independent of the jet's
$p_t$.

This last point is relevant because of the way in which fluctuations 
can affect the $A_J$ distribution.
The experimental analyses of the $A_J$ distribution select events in
which the leading jet passes some high-$p_t$ cut, say $p_t > 100\GeV$.
Events with a genuine high-$p_t$ jet are rare. There are many more
low-$p_t$ dijet events and in some small fraction of cases the
background under one of the jets may fluctuate upwards causing the
jet to pass the high-$p_t$ cut. Such events will naturally have a
large jet asymmetry, since there is no reason for the balancing jet to
also have a positive background fluctuation.
The relative contributions of different classes of events depends on
the interplay between the rareness of large background fluctuations
and the rareness of high-$p_t$ jet production as compared to low-$p_t$
jet production. 
While one can in principle estimate the potential severity
of this problem from Monte Carlo simulations, it is debatable whether
reliable enough descriptions of the PbPb medium produced at high energy exist.
Guidance from experimental measurements is therefore paramount.

One parameter that is indicative of the size of the fluctuations in the
reconstructed jet $p_t$ is their standard deviation, which we call
$\sigma_{\jet}$. 
ATLAS~\cite{ATLASTalk} has presented preliminary results for the
fluctuations from one calorimeter tower to the next. 
If scaled up by
the square-root of the number of towers in a jet (about 50 for an
$R=0.4$ jet with towers of size $0.1\times0.1$ in rapidity and
azimuth), it would suggest a value $\sigma_\jet \simeq 8.5\GeV$ for
the most central set of events.
On the other hand, the scaling of the tower fluctuations by the
square-root of the number of towers may not be a safe way of
extrapolating tower fluctuations to $\sigma_\jet$, insofar as the
background could well have local correlations (there is no clear
reason for the correlation length of such fluctuations to necessarily
be smaller than the calorimeter tower size).
Furthermore there can be other factors that contribute to a
degradation of  resolution, such as back reaction~\cite{areas} and
fluctuations in the event-by-event (or calorimeter-strip) estimation
of the background level (as discussed in sections 3.5 and A.1 of
\cite{Cacciari:2010te}).

Another way in which one may attempt to deduce the level of the
fluctuations is from a preliminary inclusive jet spectrum for
$0-100\%$ centrality (p.~41 of \cite{ATLASTalk}), which displays a
region of near Gaussian $p_t$-dependence for $p_t \lesssim 50$ that is
strongly suggestive of an origin due to fluctuations, and compatible
with $\sigma_\jet\simeq 14\GeV$. One would then expect the corresponding
$\sigma_\jet$ for $0-10\%$ central events to be somewhat larger.

Neither of the above estimates, made in the original version of this
article, is ideal. One criticism that has been made is that the tower
fluctuations do not include full calibration.
As for the inclusive jet spectrum, it mixes a genuine jet spectrum in
with the fluctuations.
However, subsequently, support for $\sigma_\jet$ values $\gtrsim
15\GeV$ has appeared in the form of a preliminary measurement of the
background fluctuations for charged-track
jets from ALICE~\cite{ALICE-resolution}. It is discussed in detail in
Appendix~\ref{sec:hydjet}.

To provide simple insight into the impact on the dijet asymmetry from
various values of $\sigma_{\jet}$, we have carried out the following
``toy'' analysis. We generate jet events with Pythia~\cite{Pythia}
(version~6.423, DW underlying-event tune~\cite{Albrow:2006rt}).
To mimic the effect of residual fluctuations following background
subtraction, we then add to the $p_t$ of each jet a random
fluctuation, generated according to a Gaussian distribution with mean
$0$ and standard deviation $\sigma_{\jet}$, independently of the jet
$p_t$ (details are given in appendix~\ref{sec:gaussian}).
These choices correspond to a perfect estimate of the average
background that needs subtracting in each event, with $\sigma_\jet$
encoding the combination of all sources of fluctuations, whether 
intrinsic fluctuations of the background, or fluctuations in the
detector's response to it.
We select events in which the leading jet has $p_t > 100\GeV$, the
subleading jet has $p_t > 25 \GeV$, both have rapidities $|y|<2.8$ and
are separated in azimuth by $|\Delta \phi| > \pi/2$ 
and for these events plot the corresponding distribution of
$A_J$, similarly to the ATLAS analysis~\cite{ATLAS:2010bu}.

\begin{figure}[t]
  \centering
  \centering
  \textsf{\large Pythia with Gaussian smearing}\\[5pt]
  \includegraphics[width=\textwidth]{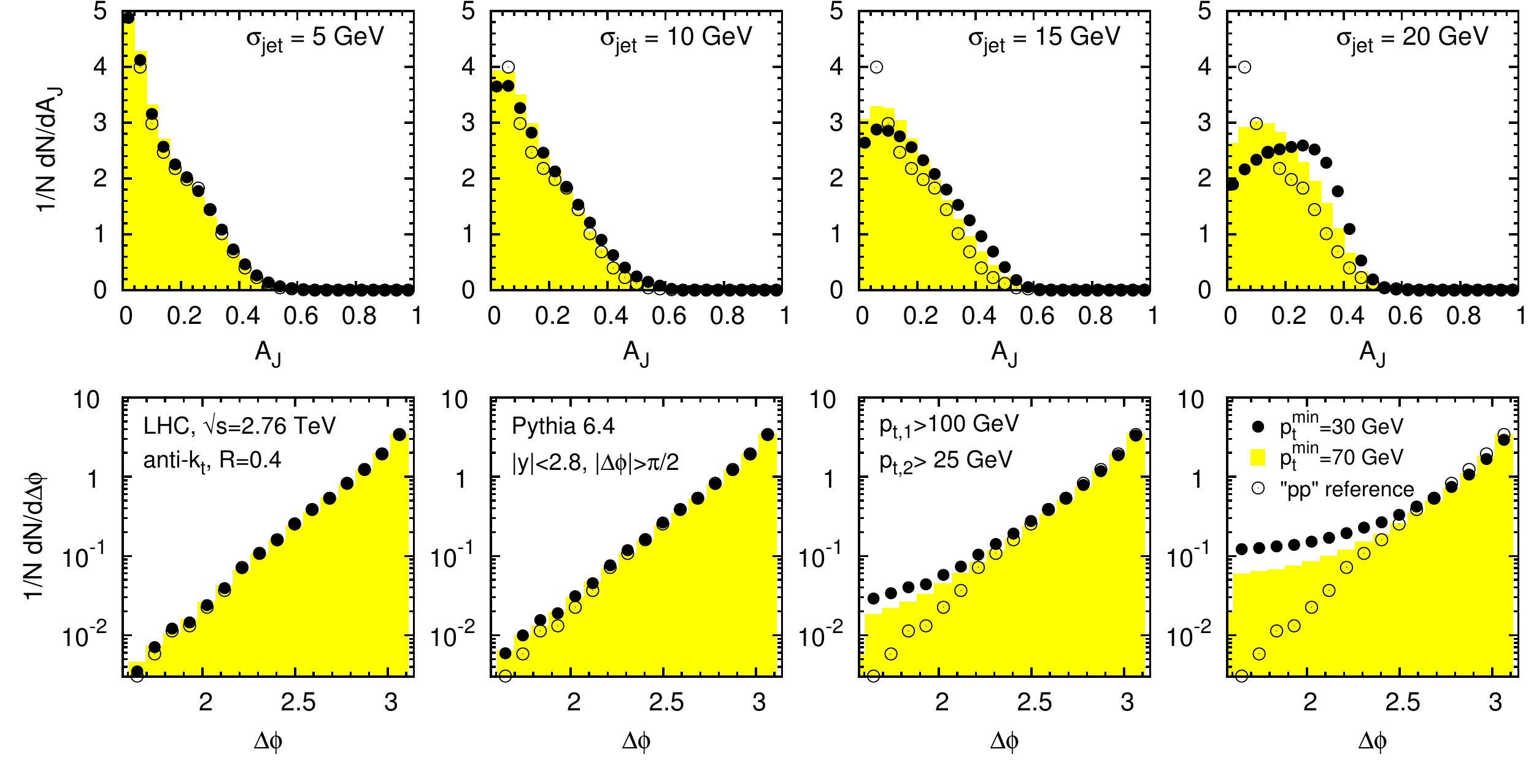}
  \caption{ %
    Simulated distribution of $A_J$ and $\Delta \phi$, as obtained
    when smearing the $p_t$ of jets from Pythia 6.4 (DW
    tune~\cite{Albrow:2006rt}) by an amount $\sigma_\jet$. 
    None of the results in this figure involved jet quenching.
    Four different $\sigma_\jet$ values are shown, and for each plot there
    are results from Pythia simulations with two different generation
    cutoffs on the $2\to2$ scattering, $p_{t}^{\min}=30\GeV$ and
    $p_{t}^{\min}=70\GeV$, so as to illustrate its impact.
    The results labelled ``pp'' reference always correspond to
    $p_{t}^{\min}=30\GeV$ with no smearing.
    Jet clustering has been performed with the anti-$k_t$
    algorithm~\cite{antikt} with $R=0.4$, as implemented in FastJet
    \cite{fastjet_fast}.  
  }
  \label{fig:our-results}
\end{figure}

The filled black points in fig.~\ref{fig:our-results} show our results
for four different values of $\sigma_{\jet}$.
One sees a clear distortion of the $A_J$ distribution as $\sigma_{\jet}$ is
increased, reminiscent of the pattern seen by ATLAS and CMS with
increasing centrality.
One key element of our simulation is that 
in generating the filled black points we chose a fairly low minimum
$p_t$ cut, $p_t^{\min}=30\GeV$, for
the underlying Pythia $2\to2$ scattering, and also verified that further 
lowering this cut made no difference for our values of $\sigma_{\jet}$.
With a larger choice, $p_t^{\min}=70\GeV$ (shaded region),\footnote{We
  understand that this was the value used in
  refs.~\cite{ATLAS:2010bu,ATLASTalk}} which would be perfectly
adequate for low $\sigma_{\jet}$, one notices that a significant part
of the effect of the background fluctuations can be missed for larger
$\sigma_{\jet}$.  This leads to the obvious implication that the
choice of $p_t^{\min}$ can play an important role, especially if
$\sigma_{\jet}$ happens to be large (or, as we have also found, if
there are significant non-Gaussianities in the
fluctuations\footnote{Significant non-Gaussianities have been observed
in~\cite{Jacobs}.}).

A complementary investigation into the impact of fluctuations can be
obtained by embedding Pythia events into
a simulated PbPb background.
A similar investigation was carried out by ATLAS, embedding events
into PbPb events as simulated by an ATLAS-specific version of HIJING~\cite{Wang:1991hta}. 
Our analysis will differ in that we study HYDJET~\cite{Lokhtin:2003ru}
rather than HIJING and use also a lower $p_t$ cutoff for the Pythia events.
The tune we use for HYDJET%
\footnote{The tuning parameters used to simulate LHC events at
  $\sqrt{s}=2.76$ TeV with HYDJET v1.6 have been extrapolated between
  the 200~GeV (RHIC) and 5.5~TeV (LHC at designed energy) values used
  in \cite{Cacciari:2010te} (footnote 7), namely $\texttt{nh}=25600$,
  $\texttt{ylfl}=3.9$, $\texttt{ytfl}=1.46$ and
  $\texttt{ptmin}=7.54\GeV$. Quenching effects have been switched off
  by setting $\texttt{nhsel}=1$.  
  The embedded events come from Pythia version 6.423, tune DW, run at
  $\sqrt{s}=2.76\TeV$.  } %
gives an average background level of $210\GeV$ per unit area for
$0-10\%$ centrality and $|\eta|<2.8$, compatible with the average jet
contamination found by ATLAS, and an average charged-particle
multiplicity for $0-10\%$ centrality of 1400 for $|\eta|<0.5$, which
is reasonably consistent with that measured by
ALICE~\cite{Aamodt:2010pb} (further comparisons are discussed in
appendix~\ref{sec:hydjet}).
HYDJET's simulation of quenching has been turned off, to avoid the
potential confusion that might arise from the quenching of hard jets
associated with the PbPb simulation rather than with the embedded
Pythia event (quenching has only a modest $5-10\%$ effect on the
HYDJET fluctuations).
Since detectors can have an impact on fluctuations, we have also
processed the events through a \mbox{simplified} calorimeter
simulation.\footnote{Charged particles with $p_t<0.5 \GeV$ are first
  removed, and the remaining particles are put on a calorimeter of
  size $0.1\times 0.1$ extending up to $|\eta|=5$ with uncorrelated
  Gaussian fluctuations of standard-deviation $0.8/\sqrt{E}$ in each
  tower and a $0\GeV$ tower threshold.
  This simple procedure gives a resolution that is $5-10\%$
  better than the true ATlAS calorimeter resolution; the comparison
  and other relevant points are discussed in appendix~\ref{sec:calo}.
  The number quoted above for the average energy flow and fluctuations
  are those obtained at calorimeter level.  }
To subtract the background from jets we have taken the area/median
method of \cite{areas,subtraction}, using, for the estimation of the
background density, a (rapidity) StripRange of half-width 0.8 centred on
the jet to be subtracted, as described in more detail in
\cite{Cacciari:2010te}.
This method should perform similarly to the ATLAS method of background
subtraction. 
With this setup, for collisions in the 0-10\% centrality range, we
find fluctuations per unit area of about $23 \GeV$ corresponding, for
anti-$k_t$ jets of radius $R=0.4$, to an expected $\sigma_{\rm jet}$
of $16\GeV$ and a measured $\sigma_{\rm jet}$ of $17\GeV$.

\begin{figure}[t]
  \centering
  \textsf{\large Pythia embedded in HYDJET}\\[5pt]
  \includegraphics[width=\textwidth]{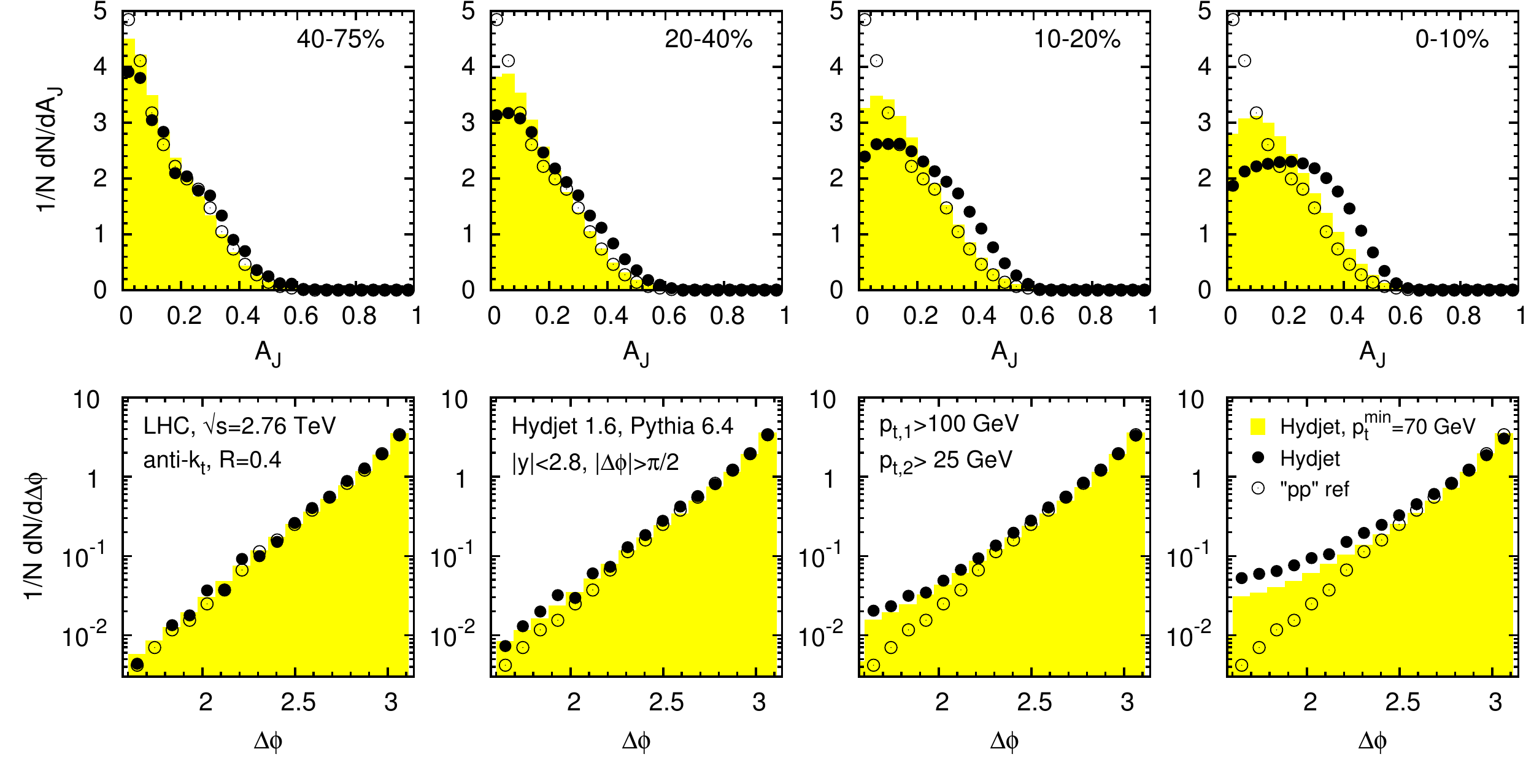}
  \caption{ %
    Simulated distribution of $A_J$ and $\Delta \phi$, as obtained
    when embedding Pythia events in a PbPb background described by
    HYDJET 1.6.
    None of the results in this figure involved jet quenching and the
    results obtained with HYDJET include a simple calorimeter simulation.
    Four different centrality regions are shown as indicated in the
    plots on the top row. For each plot there are results from Pythia
    simulations with two different generation cutoffs on the $2\to2$
    scattering, $p_{t}^{\min}=10\GeV$ and $p_{t}^{\min}=70\GeV$, so as
    to illustrate its impact.
    The results labelled ``pp'' reference always correspond to those
    of Fig. \ref{fig:our-results}.
    Jet clustering has been performed with the anti-$k_t$
    algorithm~\cite{antikt} with $R=0.4$, as implemented in FastJet
    \cite{fastjet_fast} and the heavy-ion background subtraction has
    been performed as described in \cite{Cacciari:2010te} with the
    background density estimated using a StripRange of half-width 0.8
    centred on the jet being subtracted.  }
  \label{fig:our-results-hydjet}
\end{figure}

The results we obtain from the HYDJET+Pythia simulations are presented
in Fig.~\ref{fig:our-results-hydjet} for four centrality ranges. The
empty circles labelled {\em ``pp'' reference} correspond to plain
Pythia results as for Fig. \ref{fig:our-results}. The filled black
points and shaded histogram correspond to our embedding in HYDJET events
and differ only by the $p_t^{\min}$ of the underlying Pythia $2\to2$
scattering: $10\GeV$ has been used for the former and $70\GeV$ for the
latter.\footnote{HYDJET itself generates many additional $pp$ $2\to2$
  scatterings for each heavy-ion collision, each with
  $\texttt{ptmin}=7.54\GeV$. When embedding a jet event with a
  $10\GeV$ cutoff, in most cases the two hardest jets actually
  originate from these additional HYDJET $pp$ scatterings.}

The evolution of the $A_J$ distribution with increasing centrality in
HYDJET displays a pattern similar to that observed for the Gaussian
smearing with increasing $\sigma_{\jet}$.
If anything, the distortion of the $A_J$ distribution for $0-10\%$
central HYDJET collisions is slightly more pronounced at large $A_J$ than with the
highest Gaussian smearing we used, despite the smaller $\sigma_\jet$
value from HYDJET. This could perhaps be a consequence of
non-Gaussianities in its fluctuations.
The HYDJET results also confirm the importance of the choice of
$p_{t}^{\min}$ cut on the $2\to 2$ scatters.

While the above results suggest that fluctuations could be of
relevance in interpreting the $A_J$ distributions, one should not
forget that the experiments have studied observables intended to
signal the possible presence of important effects from fluctuations.
One such observable is the fraction of energy inside a core of $R=0.2$
within the jet. A fluctuation that enhances the leading jet's $p_t$
would not necessarily be close to the centre of the jet and so should
on average reduce the core energy fraction.\footnote{This though is not entirely
  trivial, because the fluctuation may itself displace the centre of
  the jet. Furthermore any quenching of the leading jet may also
  reduce the core energy fraction.}
Preliminary data from ATLAS (p.~34 of \cite{ATLASTalk}) show a
stronger reduction in core energy fraction with increasing centrality
than in the ATLAS HIJING simulations. In our HYDJET simulations, the
core energy fraction decreases yet more rapidly, which at first sight
suggests that its fluctuations could be excessive. On the other hand,
we find that the agreement in absolute value is better for central collisions than
for peripheral collisions, complicating the
interpretation.\footnote{For the subleading jet, the centrality
  dependence is very similar for data and HYDJET, but the data are
  systematically about $0.15$ below HYDJET.}
Another cross-check on fluctuations comes from the $A_J$ distribution
for jets with $R=0.6$ (e.g. p.~48 of \cite{ATLASTalk}).
Since fluctuations should increase for a larger $R$, one would expect
them to lead to an enhancement of the high $A_J$ part of the $R=0.6$
distribution. Vacuum QCD (and jet quenching) are expected to act in
the opposite direction.
The (unquenched) HYDJET simulation shows a fairly complicated
behaviour however: the large $A_J$ ($\gtrsim 0.4$) part of the
distribution barely changes in going from $R=0.4$ to $R=0.6$, while
the distribution increases for $A_J=0.2$ (and decreases for $A_J$
near zero).
In contrast, the preliminary data decrease at large $A_J$ and, within
the (large) errors, barely change for moderate and small $A_J$,
suggesting, possibly, some non-trivial interplay between an effect
such as quenching and fluctuations.\footnote{Data from
  CMS~\cite{Collaboration:2011sx} on momentum flow in tracks, which
  appeared subsequent to the first version of this article, also
  indicate some genuine component of quenching.}

If fluctuations are relevant, as they seem to be, then it is probably
advantageous to attempt to unfold their effect, and/or to reduce their
impact by raising the jet $p_t$ thresholds.
Additionally it may be of interest to investigate methods to suppress
fluctuations (the method of \cite{Kodolova:2007hd} as used by
CMS~\cite{Collaboration:2011sx}, or
filtering/trimming/pruning~\cite{boosted_higgs,trimming,Ellis:2009me}).
Nevertheless one should be aware that such methods introduce potential
biases of their own, as has been found in earlier
studies~\cite{STAR,Cacciari:2010te}, and it then becomes important to
quantify the interplay between any quenching and the noise reduction
method.\footnote{
  Such methods discard low-momentum components of the jets, exploiting
  the fact that the background is almost entirely made of low-momentum
  particles, while for a $pp$ jet only a small fraction of its total
  momentum is contained in low-momentum components.
  In the presence of quenching, however, a larger, but unknown,
  fraction of the jet's energy may be concentrated in low-momentum
  components.
  Discarding these components is then not without risks.
  Special care should also be taken with infrared or collinear unsafe
  seeded jet algorithms, as quenching may cause a jet's high central
  energy density (the seed) to be redistributed over a broader region
  of the calorimeter.  
  Further concerns, specific to the method used by CMS, are discussed
  in appendix~\ref{sec:cms}.
}

To conclude, we have found that fluctuations can significantly affect
the dijet asymmetry $A_J$ measured in \cite{ATLAS:2010bu}.
A precise estimate of the contribution of fluctuations is therefore
important to be able to quantify the degree of quenching that is
present in the data.
A first direct estimate of these fluctuations has appeared in
preliminary form~\cite{ALICE-resolution} since the original version of
our article. 
It shows fluctuations that are well reproduced by our HYDJET
simulations and consistent also, therefore, with the upper end of the
range that we explored in the toy model.
Quantitative use of the $A_J$ data to constrain quenching
therefore probably requires that the potential bias due to fluctuations (or from
any techniques used to suppress them) be accounted for.

\section*{Acknowledgements}
\label{sec:acknowledgements}

We wish to thank Tancredi Carli, Brian Cole, Peter Jacobs, Christian
Klein-B\"osing, Michelangelo Mangano, Mateusz Ploskon, Sevil Salur,
Peter Steinberg and Urs Wiedemann, for helpful conversations, comments
and additional information.
This work was supported in part by grants ANR-09-BLAN-0060 and
PITN-GA-2010-264564.

\appendix
\section{Appendix}

Since the appearance of the first version of this article, a number of
questions have been raised concerning our analysis.
Moreover new experimental results have become available.
We address these questions here, also in light of the new data.

\subsection{Details of our Gaussian smearing procedure}
\label{sec:gaussian}
In the discussions that followed the appearance of our article, one
question that arose concerned the details of our Gaussian smearing
procedure, which we had originally omitted for brevity.
The procedure is the following: we generate Pythia events (pp,
$\sqrt{s} = 2.76\TeV$, DW tune); to these we add a low density of
``ghost'' particles ($10$ per unit area), which have infinitesimal
momentum and serve to ensure that there are no substantial empty
regions in the event; we then run the anti-$k_t$ algorithm with
$R=0.4$ on the combination of Pythia and ghost particles.
For each jet that is found (however small its momentum, and
independently of whether it involves Pythia particles or just
ghost-particles) we add to it a random $p_t$, chosen according to a
Gaussian distribution of zero mean and dispersion $\sigma_\jet$.
We then consider only the two hardest jets, requiring that they both
be above the $25\GeV$ threshold.

A criticism was made~\cite{CMS-Roland} that only ``hard'' jets should
be fluctuated and that if this choice was made then the deformation of
the $A_J$ distribution would differ substantially from that seen in
fig.~\ref{fig:our-results}. 
The criticism may have its origins in the fact that one of the LHC
experiments, CMS, uses a seeded jet algorithm. Depending on the seed
threshold, then it might indeed be the case that a hard core is a
prerequisite for a jet to be found.\footnote{CMS actually has a quite
  low seed threshold, of $1\GeV$~\cite{Collaboration:2011sx}, and so
  in practice the heavy-ion background may produce a high multiplicity
  of seeds in its own right.
  The choice of seed threshold is delicate precisely because too low a
  value brings in many ``fake'' seeds and too high a value may cause
  genuine jets to be missed, especially if quenching changes the
  relative proportion of high-$p_t$ particles in the jet.
  Additionally, seeds are intrisincally collinear unsafe.  }
For a non-seeded algorithm such as anti-$k_t$, fluctuations anywhere
in the event may cause a jet to be found, even if there was no
corresponding genuine underlying hard jet. This has to be accounted
for in the choice of original jets that receive fluctuations.
We note that had we not fluctuated zero- and low-$p_t$ jets we would
not have reproduced the broadening of the low $\Delta \phi$ tail in
fig.~\ref{fig:our-results} (which for $\sigma_\jet = 20\GeV$ is
comparable to ATLAS's results for $0-10\%$ centrality).

One criticism that would be legitimate is that some fraction of the
anti-$k_t$ jets have a very small area and should therefore be subject
to smaller fluctuations, whereas we assigned identical fluctuations to
all jets.
We have examined the impact of a modified version of our procedure, in
which the fluctuations scale in proportion to the square-root of the
jet area.
The results show slightly larger fluctuation-induced asymmetries,
perhaps because a modest fraction of anti-$k_t$ jets have areas larger
than $\pi R^2$.

\subsection{Fluctuations in ALICE data and HYDJET simulations}
\label{sec:hydjet}

To date, the most direct experimental constraint on $\sigma_{\jet}$
comes from an ALICE result presented in a talk~\cite{ALICE-resolution}
a few weeks after the appearance of our note.
This involves a measurement of the resolution in the reconstruction of
jets from charged tracks only, tested via the embedding of
single-track jets. 
The background subtraction method used there is the area/median method
of FastJet. 
We can directly compare to it by applying our analysis procedure on
the charged tracks from HYDJET, with a single additional embedded hard
track.

\begin{figure}
  \centering
  \includegraphics[height=0.7\textwidth,angle=-90]{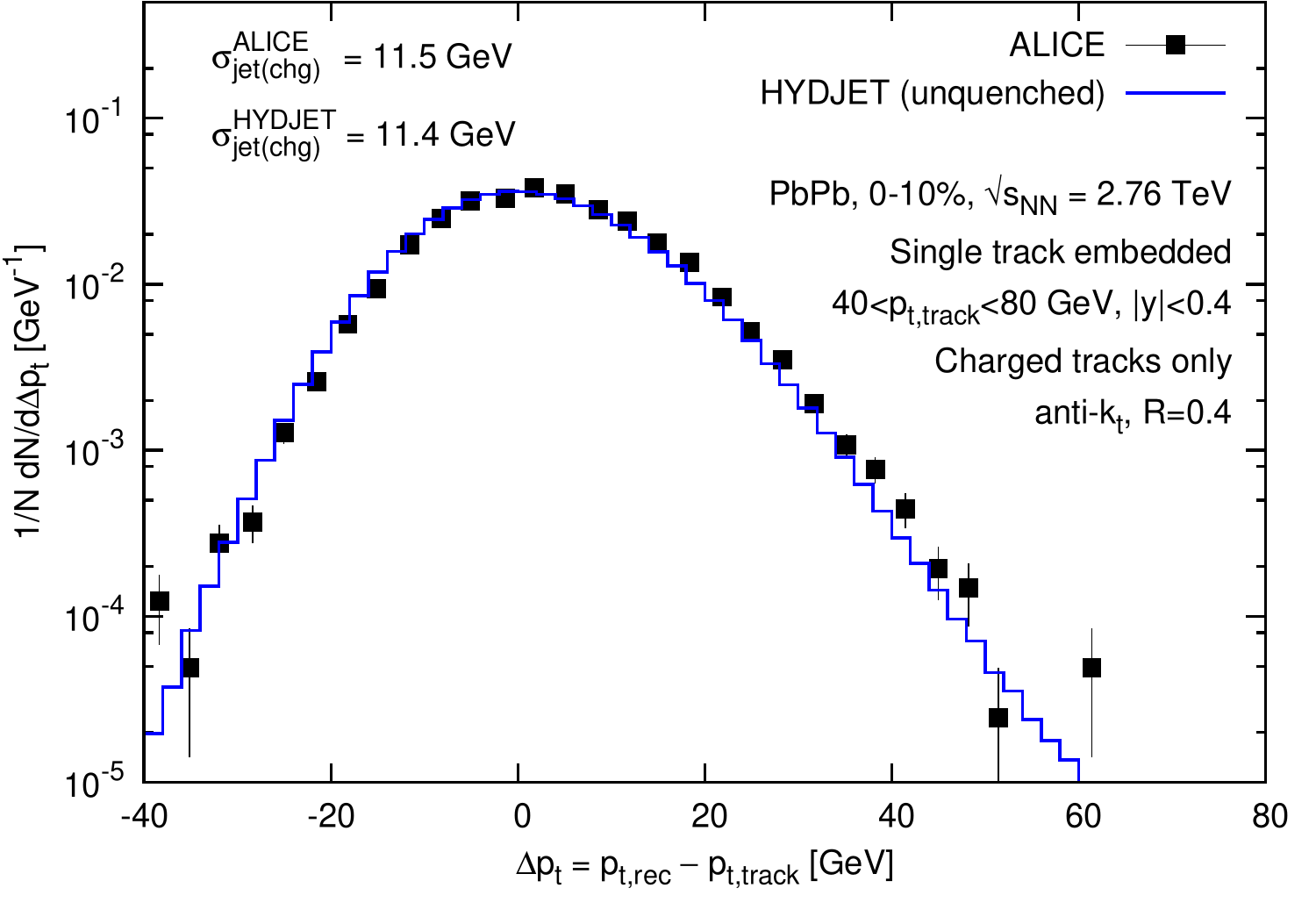}
  \caption{%
    Comparison of our HYDJET simulation to ALICE
    data~\cite{ALICE-resolution}, showing the distribution of $\Delta
    p_t = p_{t,\text{rec}} - p_{t,\text{track}}$ i.e.\ the difference
    in transverse momentum between a reconstructed charged-track jet
    (anti-$k_t$, $R=0.4$) and the single embedded charged track
    contained within it.  
    $\Delta p_t$ is calculated after 
    background subtraction with the Fastjet median/area method.  %
    The jets have been reconstructed using charged particles with
    $|\eta|<0.8$ and $p_t > 150\MeV$, whose masses have been set to
    zero by rescaling their energy.
    The background estimation in the HYDJET case used the $k_t$
    algorithm~\cite{KtAlg} with $R=0.4$, with particles and ghosts (of
    area $0.01$) up to $|y|=0.8$ and a global range up to $|y|=0.4$
    (excluding the two hardest jets within $|y|<0.8$).  %
    We understand that these choices coincide with those made by
    ALICE.  }
\label{fig:hydjet-v-alice}
\end{figure}

That comparison is given in fig.~\ref{fig:hydjet-v-alice}, which shows
the difference between the reconstructed $p_t$ of the jet that contains
the track and the $p_t$ of the track itself.
The agreement between the ALICE and HYDJET results is striking,
especially considering that our HYDJET simulation had not been directly tuned to
LHC data.
Quantitatively this can seen by comparing the $\sigma_{\jet(\chg)}$ values as
obtained from the ALICE data with those from the simulation.
We find $\sigma_{\jet\text{(chg)}} \simeq 11.5\GeV$ from the ALICE
data and $11.4\GeV$ in the HYDJET simulation.\footnote{
  There is a small residual shift between the ALICE and HYDJET
  results. When it is taken into account, the visual agreement in
  fig.~\ref{fig:hydjet-v-alice} becomes perfect. The precise origin of
  this small difference has not been identified, but small differences
  in the subtraction procedure can lead to such shifts without
  affecting the $\sigma_{\jet\text{(chg)}}$
  value~\cite{Cacciari:2010te}.
  Note that the $\sigma_{\jet\text{(chg)}}$ can, however, be affected
  by the choice of ghost area. A smaller area can reduce it by
  $0.5\GeV$. We have verified that fig.~\ref{fig:our-results-hydjet}
  remains essentially unchanged when using a smaller ghost area.
  Note that the $\sigma_{\jet\text{(chg)}}$ values quoted by ALICE
  were obtained from Gaussian fits to the $\Delta p_t < 0$
  region. Such fits give $\sigma_{\jet\text{(chg)}}$ values that are
  about $2\GeV$ smaller than results derived from the full
  distribution, which are more relevant to our studies here.  }

Several factors intervene in comparing these results to the
$\sigma_\jet$ value of $17\GeV$ discussed in the main text for full
jets embedded in HYDJET.
Firstly, in going from the embedding of a single track to that of a
full jet, an extra increase of a few percent may arise (e.g. due to
back-reaction~\cite{Cacciari:2010te}).
More importantly, assuming that a fraction $f_\chg\simeq 0.6$ of the
energy flow is carried by charged particles, one may expect an
additional factor that is somewhere between $\sim 1/\sqrt{f_\chg}
\simeq 1.3$ (taking charged and neutral components to be uncorrelated)
and $\sim 1/f_\chg \simeq 1.7$ (for $100\%$ correlated charged and
neutral components).
Finally there will be additional fluctuations from the calorimetric
nature of jet measurements.
Combining all these factors gives a result that is consistent with
the $\sigma_\jet \simeq 17\GeV$ quoted in the main text.

Ultimately therefore, the ALICE results support the choices that we
made in estimating the possible order of magnitude of
fluctuation-induced contributions to the jet-asymmetry
measurement. Note however that other issues also need to be taken in
to account at the $1-2 \GeV$ level (for example detector effects). An
analysis at this level of accuracy can therefore probably only be
performed in conjunction with a fu ll detector simulation.

\subsection{Suitability of calorimeter simulation}
\label{sec:calo}

\begin{figure}
  \centering
  \includegraphics[width=0.7\textwidth]{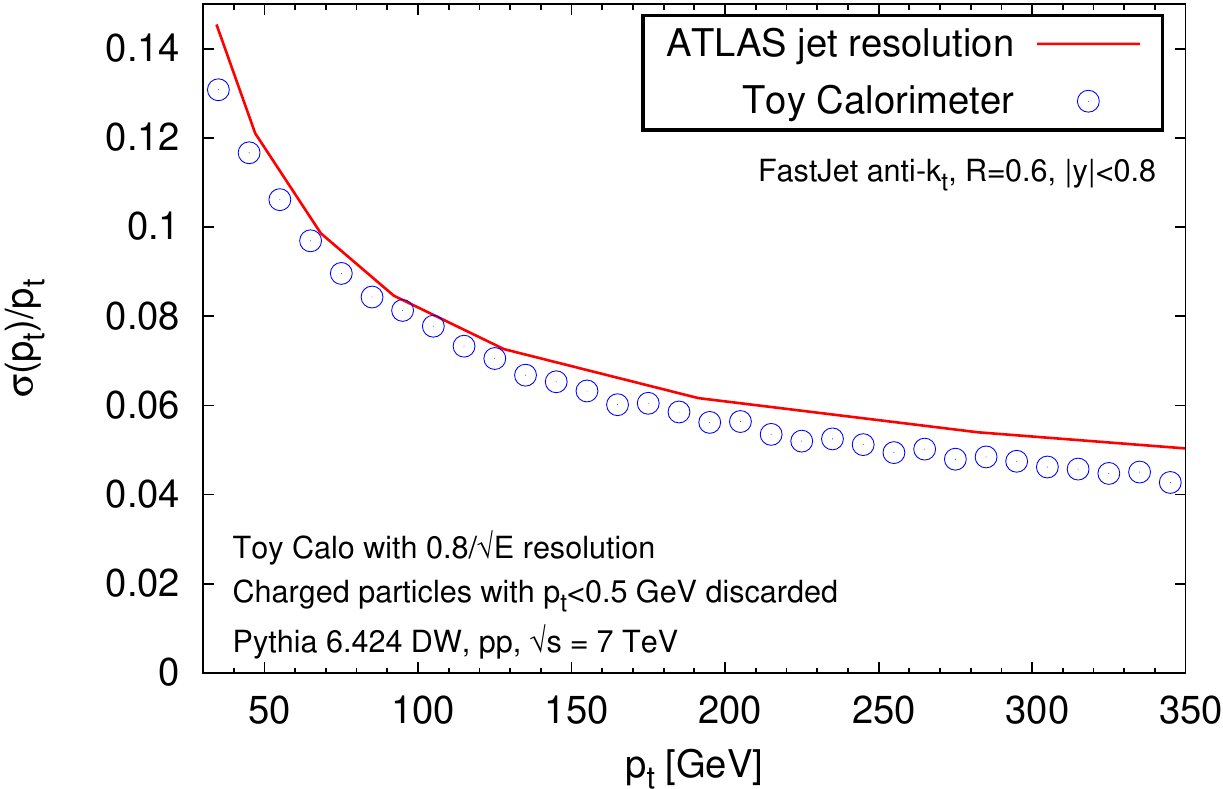}
  \caption{Comparison of jet resolution obtained with our toy
    calorimeter to the measured ATLAS jet resolution~\cite{ATLAS-resolution}.}
  \label{fig:calo-validation}
\end{figure}

One objection that has been raised concerning our HYDJET results is
that our toy calorimeter simulation suffered from an overly
pessimistic resolution.
To obtain a detailed description of jet response, ideally many effects
need to be taken into account, including tower thresholds, different
kinds of noise term that are independence of energy, scale as
$1/\sqrt{E}$ and as $1/E$, different responses to photons and 
charged hadrons, the effect of magnetic fields, and the degradations
of resolution due to detector elements that lie between the interaction
point and the calorimeter.
A full simulation of all these effects goes beyond the scope of this
article, therefore we made a simple approximation involving towers of
size $0.1\times 0.1$ with a resolution of $0.8/\sqrt{E}$, together
with the removal of charged tracks with $p_t < 0.5\GeV$ (on the
grounds that they would be bent away from the calorimeter by the
magnetic field).
The $0.8/\sqrt{E}$ term is larger than the corresponding term for the
ATLAS hadronic calorimeter ($\sim 0.5/\sqrt{E}$), however this
difference is justified by the fact that we neglect many other sources
of detector fluctuation.
The validation of this statement is given through
fig.~\ref{fig:calo-validation}, which compares the jet resolution that
we have observed with our calorimeter (matching calorimeter jets
within $\Delta R < 0.25$ of the two hardest particle-level jets in an
event) to that found by ATLAS.
One sees that rather than overestimating calorimeter fluctuations, we
actually underestimate them by $5-10\%$ over the full range of $p_t$ in
which jet resolution measurements exist.
Note also that ATLAS has shown that its calorimeter response is
similar for jets and heavy-ion backgrounds (p.~37 of
\cite{ATLASTalk}).

A further potential issue comes from the fact that an experiment's
magnetic field, by bending particles differently according their
$p_t$, may smear any angular correlations that are present in the
background fluctuations.
Our calorimeter simulation did not by default consider the impact of a
magnetic field (other than by removing charged particles with $p_t <
0.5\GeV$).
We have, however, verified that if one accounts for the effect of a
magnetic field on the charged-particles' azimuth when they reach the
calorimeter, the FastJet median/area estimate of fluctuations is
reduced by an amount of order $1\%$.
This value was obtained with a configuration similar to that used by
ATLAS, i.e. a longitudinal magnetic field of strength $B=2$~T, with a
calorimeter at a radius of $1.5$~m from the beam.
This result gives us confidence that our original approximation was
not unreasonable.

If is to be noted that magnetic fields could also have an impact on
measurements of quenched jets (independently of background-related
issues), insofar as quenching may alter the relative fractions of
different momentum components within a jet and therefore also give an
appearance of additional jet broadening.

\subsection{The CMS jet-reconstruction procedure}
\label{sec:cms}

It was not the purpose of this article to discuss the CMS
results~\cite{Collaboration:2011sx}, insofar as they appeared
subsequent to the first version of this article.
However, certain criticisms of our work have been made based on
results presented by CMS, and so deserve comment.

An objection that has been raised is that CMS explicitly show jet
resolution results that are incompatible with the values we have
investigated.
For example, fig.~4f of \cite{Collaboration:2011sx} indicates that
$40\GeV$ jets ($0-10\%$ centrality) have about $20\%$ resolution,
i.e. a $\sigma_\jet \simeq 8 \GeV$, well below the largest values that
we have discussed here.
In this context, however, it is important to be aware that CMS's jet
reconstruction procedure differs substantially from ATLAS's. 
One of the differences is that the method used to subtract the
heavy-ion background involves a noise reduction
technique~\cite{Kodolova:2007hd} that estimates the mean $\mu$ and
standard deviation $\sigma_\tower$ of the calorimeter towers' $p_t$ deposits (after
excluding hard jets) and then subtracts $\mu + \sigma_\tower$ from each
tower's $p_t$.
Only towers that are positive are then retained.

To understand the potential benefits and biases of this method, let us
make the assumption that $\mu$ and $\sigma_\tower$ are well determined
(there are a number of complications in their practical determination,
e.g.\ with respect to the exclusion of hard jets, which may degrade
the performance of the method with respect to the analysis that
follows).
Let us also assume purely Gaussian noise (again, this is probably
optimistic). Then the fraction of towers retained is
\begin{equation}
  \int_{\sigma_\tower}^\infty dx \frac{1}{\sqrt{2\pi} \sigma_\tower}
  e^{-\frac{x^2}{2\sigma_\tower^2}} 
  \simeq 0.1587\,,
\end{equation}
where $x = p_{t,\tower} - \mu$. 
Since all of these towers are positive, they induce a systematic
offset in the overall reconstructed jet momentum
\begin{equation}
  \label{eq:deltapt-noise}
  \langle \delta p_{t,\jet}^{\text{noise}} \rangle
  = N_\tower \langle \delta p_{t,\tower}^{\text{noise}} \rangle
  %
  = N_\tower \int_{\sigma_\tower}^\infty dx \frac{(x -
    \sigma_\tower)}{\sqrt{2\pi} \sigma_\tower} 
  e^{-\frac{x^2}{2\sigma_\tower^2}} 
  \simeq 0.0833 \, \sigma_\tower \, N_\tower\,,
\end{equation}
where $N_\tower$ is the total number of towers that are contained in a
jet ($\pi R^2$ divided by the tower area, which is $0.087\times
0.087$, i.e.\ $N_\tower \simeq 104$ with $R=0.5$ as used by CMS).
For $\sigma_\tower \simeq 1-2\GeV$, this would correspond to an $8-16\GeV$ bias.

The residual fluctuations after the noise suppression can be estimated
as 
\begin{subequations}
  \label{eq:sigmajet-noise-suppressed}
  \begin{align}
    (\sigma_{\jet}^\text{noise-suppressed} )^2 &= N_\tower \left[
      \langle (\delta p_{t,\tower}^{\text{noise}})^2\rangle - \langle
      \delta p_{t,\tower}^{\text{noise}}\rangle^2 \right]
    \\
    &= N_\tower \left[ \int_{\sigma_\tower}^\infty dx \frac{(x -
        \sigma_\tower)^2}{\sqrt{2\pi} \sigma_\tower}
      e^{-\frac{x^2}{2\sigma_\tower^2}} - \langle \delta
      p_{t,\tower}^{\text{noise}}\rangle^2 \right]
    \\
    &\simeq (0.262\, \sigma_\tower)^2\, N_\tower\,,
  \end{align}
\end{subequations}
i.e.\ a $74\%$ reduction in the noise as compared to subtraction
procedures without noise suppression, which just give $\sigma_{\jet}^2
\simeq \sigma_\tower^2\, N_\tower$.
This is a potential strength of the CMS noise-reduction technique,
but it comes at the price of introducing a significant $p_t$
offset, eq.~(\ref{eq:deltapt-noise}).
Since noise-reduction has such a large impact on the fluctuations it
is not possible to draw any conclusion about ATLAS results based on
the quoted CMS jet resolution after noise-suppression.

In practice, in simulations that embed hard $pp$ jets in a
heavy-background, the offset of eq.~(\ref{eq:deltapt-noise}) will not
be directly seen.
This is because embedded hard jets are not just pure noise, but
involve some number of towers that are far above the $\mu +
\sigma_\tower$ threshold.
These towers will all be retained.
However, relative to the original $pp$ towers, there will now be an
offset of $-\sigma_\tower$ for each of these towers. 
Defining the the calorimeter ``occupancy'' of a normal $pp$ jet to be
$f$ (i.e. a $pp$ jet contains on average $f N_\tower$ active towers),
then the total offset from this effect will be 
\begin{equation}
  \langle \delta p_{t,\jet}^{\text{hard}} \rangle \simeq
  - f N_\tower \sigma_\tower\,.
\end{equation}
In the the limit in which $f$ is small its impact can be neglected in 
eqs.~(\ref{eq:deltapt-noise},\ref{eq:sigmajet-noise-suppressed}).
For hard QCD vacuum jets we find that it has a value $f \simeq
0.1$.\footnote{ It is not clear however whether this is truly small
  enough for its quantitative impact in
  eqs.~(\ref{eq:deltapt-noise},\ref{eq:sigmajet-noise-suppressed}). to
  be entirely negligible.}

Taking into account this effect and the offset of
eq.~(\ref{eq:deltapt-noise}), gives us an overall offset of 
\begin{equation}
  \langle \delta p_{t,\jet}^{\text{overall}} \rangle 
  = 
  \langle \delta p_{t,\jet}^{\text{noise}}\rangle + \langle \delta
  p_{t,\jet}^{\text{hard}} \rangle
  \simeq
  (0.0833 - f) N_\tower \sigma_\tower\,.
\end{equation}
Thus the net bias is small, but only due to a fortuitous cancellation
between two effects with very different physical origins.\footnote{The
  cancellation may be less fortuitous that it seems, since the choice
  to subtract $\mu+x \sigma_\text{tower}$ with $x=1$ is a priori
  arbitrary and may have been tuned specifically to some subset of
  vacuum QCD jets. Note also that fluctuations in $f$ from one jet to
  another can be substantial and this can partially counteract the
  reduction in $\sigma_\jet$ due to the noise supression of the
  background fluctuations.}
Both of these effects are at the same $10\%$ level as the overall
impact of quenching (e.g.\ fig.~12 of \cite{Collaboration:2011sx}).
Furthermore, it is not immediately clear that $f$ for quenched jets
is the same as 
for pp jets and one may even expect it to be larger, especially for the
away-side jet, leading to an additional negative offset for that jet,
thereby artificially enhancing the asymmetry.
Therefore any quantitative analysis of quenching in heavy-ion
collisions that relies on noise reduction should also perform an
analysis of systematic errors due to any biases associated with
potential medium-induced modifications of $f$.

The discussion also implies that it is difficult to relate the large
number of supporting plots by CMS (e.g.\ track-properties of
calorimeter jets) directly to the ATLAS case.

\end{document}